\documentclass[prl,twocolumn,showpacs]{revtex4}
\usepackage{graphicx,epsf}

\newcommand{\bq}{{\bf q}}
\newcommand{\bk}{{\bf k}}

\newcommand{\br}{{\bf r}}

\newcommand{\ptilde}{\tilde{p}}
\newcommand{\stilde}{\tilde{s}}
\newcommand{\ctilde}{\tilde{c}}
\newcommand{\ltilde}{\tilde{l}}

\newcommand{\rhobar}{\bar{\rho}}

\newcommand{\nubar}{\bar{\nu}}
\newcommand{\rhobarbar}{\bar{\bar{\rho}}}

\newcommand{\Hhat}{\hat{H}}

\begin{document}

\def\tende#1{\,\vtop{\ialign{##\crcr\rightarrowfill\crcr
\noalign{\kern-1pt\nointerlineskip}
\hskip3.pt${\scriptstyle #1}$\hskip3.pt\crcr}}\,}

\title{Possible Reentrance of the Fractional Quantum Hall Effect in
  the Lowest Landau Level}
\author{M.\ O.\ Goerbig$^{1,2}$, P.\ Lederer$^2$, and  
C.\ Morais\ Smith$^{1,3}$}

\affiliation{$^1$D\'epartement de Physique, Universit\'e de Fribourg, P\'erolles,  CH-1700 Fribourg, Switzerland.\\
$^2$Laboratoire de Physique des Solides, Bat.\,510, UPS (associ\'e au CNRS), F-91405 Orsay cedex, France.\\
$^3$Institute for Theoretical Physics, Utrecht University, Leuvenlaan 4,
3584 CE Utrecht, The Netherlands.}

\begin{abstract}
In the framework of a recently developed model of interacting composite 
fermions, we calculate the energy of different solid and Laughlin-type liquid
phases of spin-polarized composite fermions. The liquid phases have a
lower energy than the
competing solids around the electronic filling factors $\nu=4/11,6/17$, and 
$4/19$ and may thus be responsible for the fractional quantum Hall 
effect at $\nu=4/11$. The alternation between solid and liquid phases 
when varying the magnetic field may 
lead to reentrance phenomena in analogy with the observed reentrant integral 
quantum Hall effect. 
\end{abstract}
\maketitle

Recent experiments by Eisenstein {\sl et al.} have revealed a
reentrance of the integer quantum Hall effect (IQHE) when the two spin
branches of the lowest Landau level (LL) are completely filled and the
first excited LL ($n=1$) starts to be populated \cite{exp3}. This
phenomenon arises when the partial filling of the topmost LL is varied,
and it is due to an alternation of electron-solid phases, such as
the Wigner crystal (WC) or a bubble crystal, 
and quantum liquids, which display the fractional
quantum Hall effect (FQHE) \cite{goerbig03}. 
Both liquid and solid phases have their origin in the
Coulomb interaction between the electrons, which is the relevant energy
scale when the last occupied LL is not completely filled. In this
case, intra-LL excitations coupling states with the same kinetic
energy are possible, in contrast to the case of completely filled LLs, where 
such excitations are prohibited by the Pauli principle. Electrons in
the completely filled lower LLs may then be considered as an inert
homogeneous background. 

The FQHE in the lowest LL occurs at the filling factors 
$\nu=n_{el}/n_B=p/(2ps+1)$, where $n_{el}$ denotes the electronic 
and $n_B=eB/h$ the flux density. It may be interpreted
as an IQHE of a new type of particle, a so-called composite fermion
(CF) \cite{jain}, which consists of an electron and a vortex-like object,
which carries $s$ pairs of flux quanta. Because of its reduced
coupling $(eB)^*=eB/(2ps+1)$ to the magnetic field $B$, $p$ CF-LLs are
completely filled when $\nu^*=p$, where the CF-LL filling $\nu^*$ factor is 
related to the electronic one by $\nu=\nu^*/(2s\nu^*+1)$. 
The similarity between the IQHE and the FQHE naturally raises the question
whether a phenomenon analogous to the reentrant IQHE may occur for CFs
when a higher CF-LL ($p\geq1$) is partially filled. In this case,
residual interactions between CFs may lead to the
formation of CF-solid phases, such as stripes and bubbles 
\cite{leescjain}, or to incompressible liquids, which may be
interpreted in terms of a second generation of CFs (C$^2$Fs)
\cite{goerbig04,lopezfradkin,pan}. The existence of an incompressible
CF liquid is supported by the recent observation of a FQHE at
$\nu=4/11$, which corresponds to a CF filling factor $\nu^*=1+1/3$, by
Pan {\sl et al.} \cite{pan}. Such a state had been conjectured
before by Mani and v.\ Klitzing based on arguments concerning the
self-similarity of the Hall-resistance curve \cite{mani}.

In this Letter, we investigate several spin-polarized CF-solid and liquid
states in a recently developed model of interacting CFs in a partially
filled CF-LL \cite{goerbig04}, which incorporates the self-similarity of
the FQHE \cite{mani} by an iterative projection of the dynamics to a single
CF-LL. The form of the CF interaction has been derived in
the framework of the Hamiltonian theory, proposed by Murthy and Shankar 
\cite{MS}. The basic assumptions are that the degenerate CF-LLs remain stable
in the presence of the residual CF interactions and that the low-energy degrees
of freedom are described by intra-level excitations, in agreement with 
complementary studies in the wave-function approach 
\cite{leescjain,mandaljain,changjain}.
The energies of the competing phases
are calculated directly in the thermodynamic limit. In contrast to
prior investigations by Lee {\sl et al.} 
\cite{leescjain}, 
our energy calculations suggest the possibility of incompressible 
Laughlin-type liquid states of CFs. Such states have a lower energy than CF
solids at certain filling factors also in higher CF-LLs ($p\geq 1$),
{\sl e.g.} at $\nu=4/11,6/17,4/19$, and $11/27$. An alternation between
these liquid and CF-solid phases with varying CF filling occurs in our energy 
studies and may lead to observable reentrance phenomena in the FQHE regime. 

The model Hamiltonian of CFs, the dynamics of which are restricted to the
$p$-th CF-LL, is given by \cite{goerbig04}
\begin{equation}
\label{equ01}
\Hhat(s,p)=\frac{1}{2A}\sum_{\bq}v_{s,p}^{CF}(q)\rhobar^{CF}(-\bq)\rhobar^{CF}(\bq),
\end{equation}
with the measure $\sum_{\bq}=A\int d^2q/(2\pi)^2$ in terms of the
total area $A$ and the CF-interaction 
potential ($l_B\equiv1$)
\begin{eqnarray}
\label{equ02}
\nonumber
v_{s,p}^{CF}(q)&=&\frac{2\pi
  e^2}{\epsilon q}e^{-q^2l_B^{*2}/2}\left[L_p\left(\frac{q^2l_B^{*2}c^2}{2}\right)
\right.\\ &&\left.
-c^2e^{-q^2/2c^2}L_p\left(\frac{q^2l_B^{*2}}{2c^2}\right)\right]^2,
\end{eqnarray}
where $c^2=2ps/(2ps+1)$ is the vortex charge, $l_B^*=1/\sqrt{1-c^2}$
is the CF magnetic length, and $L_p(x)$ denotes a Laguerre polynomial
\cite{MS}. The potential takes into account the form of CFs in the $p$-th 
CF-LL. The projected CF-density operators satisfy the
Girvin-MacDonald-Platzman algebra \cite{GMP},
$[\rhobar^{CF}(\bq),\rhobar^{CF}(\bk)]=2i\sin\left[(\bq\times\bk)_zl_B^{*2}/
2\right]\rhobar^{CF}(\bq+\bk)$,
if one replaces the electronic by the CF magnetic length. This model of 
interacting CFs is similar to the one of interacting electrons in a single
LL, and this property allows for the use of the standard theoretical 
approaches to describe the different phases. Note that we consider an ideal
two-dimensional electron gas. Finite-width and impurity effects are not
taken into account. In contrast to Ref. \cite{goerbig04}, where the activation
gaps of C$^2$F states have been discussed, we neglect inter-CF-LL mixing, which
may in principle be included in a dielectric function and which screens the CF 
interaction potential \cite{goerbig04}.

The energies of the different CF-solid phases are calculated in the 
Hartree-Fock approximation (HFA), in analogy with the electron-solid phases
in higher LLs \cite{FKS,moessner,goerbig03}. We concentrate on triangular 
bubble crystals with $M$ CFs per bubble and CF-stripe phases, 
which become relevant at half-filled levels. The CF-WC is formally treated as
a bubble crystal with $M=1$. The Hamiltonian (\ref{equ01}) may be rewritten
in the HFA,
$$\hat{H}_{HF}(s,p)=\frac{1}{2A}\sum_{\bf q}u_{s,p}^{HF}(q)\langle \rhobar^{CF}(-{\bf q})\rangle \rhobar^{CF}({\bf q}),$$
with the Hartree-Fock potential
$$u_{s,p}^{HF}(q)=v_{s,p}^{CF}(q)-\frac{1}{n_B^*A}\sum_{\bf p}v_{s,p}^{CF}(p)e^{-i(p_xq_y-p_yq_x)l_B^{*2}},$$
where $n_B^*=1/2\pi l_B^{*2}$ is the renormalized flux density.
The second term takes into account quantum-mechanical exchange
effects. The average $\langle \rhobar^{CF}(\bq)\rangle$ plays the role
of an order parameter, 
$\Delta^*(\bq)=\langle \rhobar^{CF}(\bq)\rangle/n_B^*A=\int d^2r\nubar^*(\br)e^{i\bq\cdot\br}/A,$
and is related to the {\sl local} CF filling factor $\nubar^*(\br)$ of the 
$p$-th CF-LL by Fourier transformation. In terms of the order parameter, the
cohesive energy $E_{coh}=\langle \hat{H}_{HF}(s,p)\rangle/\nubar^*n_B^*A$
of the CF-solid phases, with the partial CF filling factor 
$\nubar^*=A^{-1}\int d^2r\nubar^*(\br)$, is
\begin{equation}
E_{coh}^{CF-solid}(s,p;\nubar^*)=\frac{n_B^*}{2\nubar^*}\sum_{\bq}
u_{s,p}^{HF}(q)|\Delta^*(\bq)|^2.
\end{equation}
The order parameter for the CF-bubble crystal is given in terms of the
Bessel function $J_1(x)$,
$$\Delta_B^*(\bq)=\frac{2\pi\sqrt{2M}l_B^*}{Aq}J_1(q\sqrt{2M}l_B^*)\sum_j e^{i\bq\cdot{\bf R}_j},$$
where ${\bf R}_j$ are the lattice vectors of the triangular lattice with a
lattice spacing $\Lambda_B=(4\pi M/\sqrt{3}\nubar^*)^{1/2}l_B^*$. One thus 
obtains for the cohesive energy of the CF-bubble crystal
{\small
\begin{equation}
\label{equ03}
E_{coh}^{CF-B}(s,p,M;\nubar^*)=\frac{n_B^*\nubar^*}{M}\sum_{{\bf G}_l\neq 0} 
u_{s,p}^{HF}({\bf G}_l)
\frac{J_1^2(\sqrt{2M}|{\bf G}_l|l_B^*)}{|{\bf G}_l|^2l_B^{*2}},
\end{equation}}
where ${\bf G}_l$ are the vectors of the reciprocal lattice. The order 
parameter of the CF-stripe phase,
$$\Delta_S^*(\bq)=\frac{2}{L_x}\delta_{q_y,0}\frac{\sin\left(q_x\Lambda_S\nubar^*/2\right)}{q_x}\sum_je^{iq_xj\Lambda_S},$$
with the system extension $L_x$ in the $x$-direction, yields the cohesive 
energy
{\small 
\begin{equation}
\label{equ04}
E_{coh}^{CF-S}(s,p,\Lambda_S;\nubar^*)=\frac{n_B^*}{2\pi^2\nubar^*}
\sum_{j\neq0}u_{s,p}^{HF}\left(q=\frac{2\pi}{\Lambda_S}j\right)
\frac{\sin^2(\pi\nubar^* j)}{j^2}.
\end{equation}}
The stripe periodicity $\Lambda_S$ is a variational parameter with respect to 
which the energy is minimized. It scales with the CF cyclotron radius 
$R_C^*=l_B^*\sqrt{2p+1}$, and one finds $\Lambda_S(s=1,p=1)=1.95 R_C^*$, 
$\Lambda_S(s=1,p=2)=1.8R_C^*$, and $\Lambda_S(s=2,p=1)=1.9 R_C^*$ for
the optimal stripe periodicity at $\nubar^*=1/2$.

The cohesive energy of the CF solids has to be compared to the energy of
Laughlin-type quantum liquids, which may occur around the ``magical'' filling
factors $\nubar^*=1/(2\stilde+1)$, with integral $\stilde$. Because of their
strong correlations, these liquid phases cannot be treated in the HFA, and
one has to use Laughlin's wave functions \cite{laughlin}, generalized 
to an arbitrary LL \cite{macdonald1}. Their cohesive energy is given by 
\begin{equation}
E_{coh}^L(s,p;\stilde)=\frac{\nubar^*}{\pi}\sum_{m=0}^{\infty}c_{2m+1}^{\stilde}V_{2m+1}(s,p),
\end{equation}
with Haldane's pseudopotentials \cite{haldane}
$V_{2m+1}(s,p)=(2\pi/A)\sum_{\bq}v_{s,p}^{CF}(q)L_{2m+1}(q^2l_B^{*2})\exp(-q^2l_B^{*2}/2)$, and 
the expansion coefficients $c_{2m+1}^{\stilde}$ characterize the wave 
function. In contrast to Ref.\ \cite{leescjain}, where a few number of 
pseudopotentials, which have been determined numerically, were used to 
construct a CF interaction potential, here, 
they are obtained to arbitrary order from the analytical expression of the 
CF interaction potential (\ref{equ02}). As pointed out by the authors of
Ref.\ \cite{leescjain}, the construction of an interaction potential from
pseudopotentials is not unique. The expansion coefficients 
$c_{2m+1}^{\stilde}$ are 
derived from sum rules \cite{GMP,girvin}, which are considered as a set of 
linear equations \cite{goerbig01}. This method
yields results deviating less than $1\%$ from numerical studies and is thus 
sufficiently accurate for the present investigations. 

Away from the magical filling factors, the energy of the CF-liquid phases is 
raised by the excited quasi-particles [for $\nubar^*>1/(2\stilde+1)$]
or quasi-holes [for $\nubar^*<1/(2\stilde+1)$], which are separated by
a gap from the incompressible liquid state. They may be interpreted as
C$^2$Fs or C$^2$F holes 
promoted to the next higher C$^2$F level \cite{goerbig04,lopezfradkin}. 
Their energy is calculated analytically in the framework of the Hamiltonian
theory \cite{MS}. One finds 
\begin{eqnarray}
\nonumber
\Delta_{s,p}^{qp}(\stilde,\ptilde)&=&\frac{1}{2A}\sum_{\bq}v_{s,p}^{CF}(q)\langle \ptilde|\rhobarbar(-\bq)\rhobarbar(\bq)|\ptilde \rangle\\
\nonumber
&&-\frac{1}{A}\sum_{\bq}v_{s,p}^{CF}(q)\sum_{j'=0}^{\ptilde -1}|\langle \ptilde|\rhobarbar(\bq)|j'\rangle|^2
\end{eqnarray}
for the quasi-particle energies and 
\begin{eqnarray}
\nonumber
\Delta_{s,p}^{qh}(\stilde,\ptilde)&=&-\frac{1}{2A}\sum_{\bq}v_{s,p}^{CF}(q)\langle \ptilde-1|\rhobarbar(-\bq)\rhobarbar(\bq)|\ptilde-1 \rangle\\
\nonumber
&&+\frac{1}{A}\sum_{\bq}v_{s,p}^{CF}(q)\sum_{j'=0}^{\ptilde-1}|\langle \ptilde-1|\rhobarbar(\bq)|j'\rangle|^2
\end{eqnarray}
for the quasi-hole energies, where the matrix elements ($j\geq j'$) are given 
by
{\small
\begin{eqnarray}
\nonumber
\langle j|\rhobarbar(\bq)|j'\rangle=\sqrt{\frac{j'!}{j!}}\left(\frac{-i(q_x-iq_y)\ltilde\ctilde}{\sqrt{2}}\right)^{j-j'}e^{-q^2\ltilde^2\ctilde^2/4}\\
\nonumber
\times\left[L_{j'}^{j-j'}\left(\frac{q^2\ltilde^2\ctilde^2}{2}\right)-\ctilde^{2(1-j+j')}e^{-q^2/2\ctilde^2}L_{j'}^{j-j'}\left(\frac{q^2\ltilde^2}{2\ctilde^2}\right)\right],
\end{eqnarray}}
with the C$^2$F magnetic length $\ltilde=l_B^*/\sqrt{1-\ctilde^2}$, in 
terms of the C$^2$F-vortex charge 
$\ctilde^2=2\ptilde\stilde/(2\ptilde\stilde+1)$ \cite{goerbig04}. Here, we 
are interested only in states from the Laughlin series ($\ptilde=1$), and one 
finds 

\vspace{0.2cm}
\noindent
{\small
\begin{tabular}{|c||c|c||c|c|}
\hline
~ & $\Delta^{qp}(\stilde=1)$ & $\Delta^{qp}(\stilde=2)$ &
$\Delta^{qh}(\stilde=1)$ & $\Delta^{qh}(\stilde=2)$ \\ \hline \hline
$s=1$, $p=1$ & 0.04172 & 0.03065 & -0.01567 & -0.01309\\ \hline
$s=1$, $p=2$ & 0.02330 & 0.01958 & -0.01277 & -0.00967\\ \hline
$s=2$, $p=1$ & 0.01727 & 0.01391 & -0.00859 & -0.00672\\ \hline
\end{tabular}}
\vspace{0.2cm}

\noindent
in units of $e^2/\epsilon l_B$. The cohesive energy of the CF-liquid phases
thus becomes
\begin{eqnarray}
\label{equ05}
E_{coh}^{CF-liq}(s,p;\stilde,\nubar^*)&=&E_{coh}^{L}(s,p;\stilde)\\
\nonumber
&&+[\nubar^*(2\stilde+1)-1]\Delta_{s,p}^{qp/qh}(\stilde,\ptilde=1),
\end{eqnarray}
where the residual C$^2$F interactions have been neglected. This approximation
is valid in the vicinity of the magical filling factors, {\sl i.e.} at low
C$^2$F density. The complete energy curve for the quantum-liquid phases
would require a full account of these interactions, which is beyond
the scope of the present studies.

\begin{figure}
\epsfysize+5.4cm
\epsffile{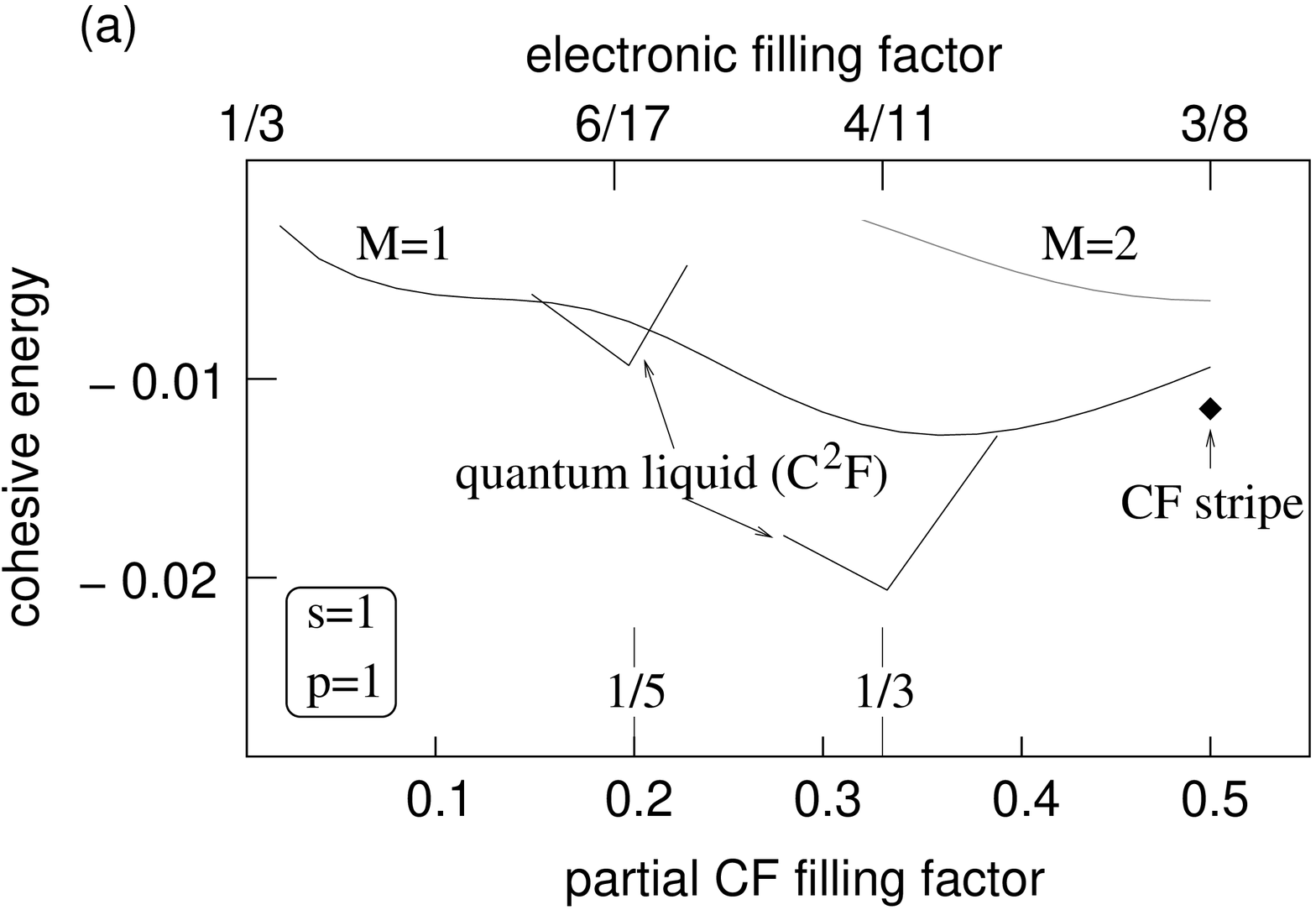}
\epsfysize+5.2cm
\epsffile{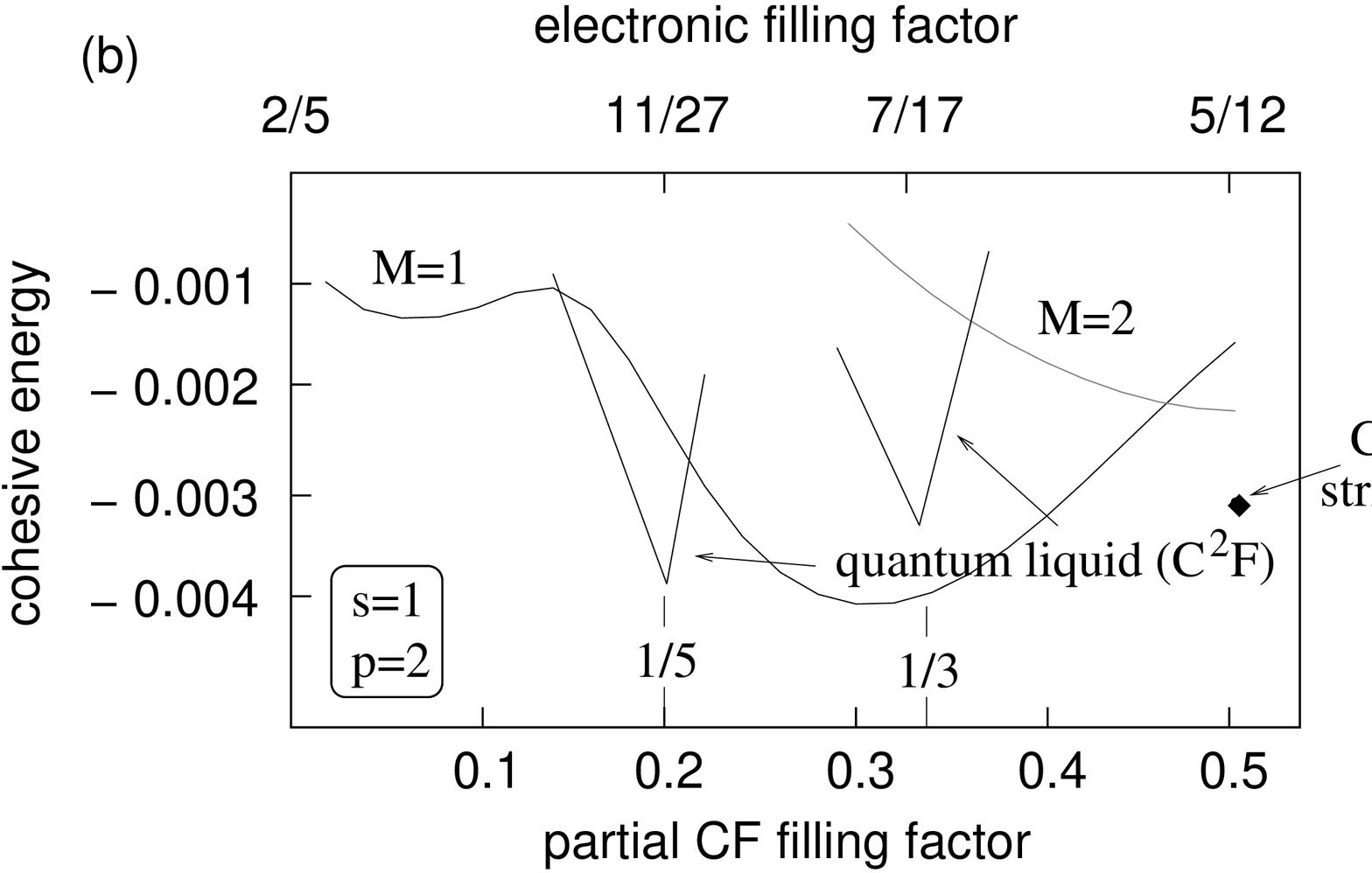}
\epsfysize+5.4cm
\epsffile{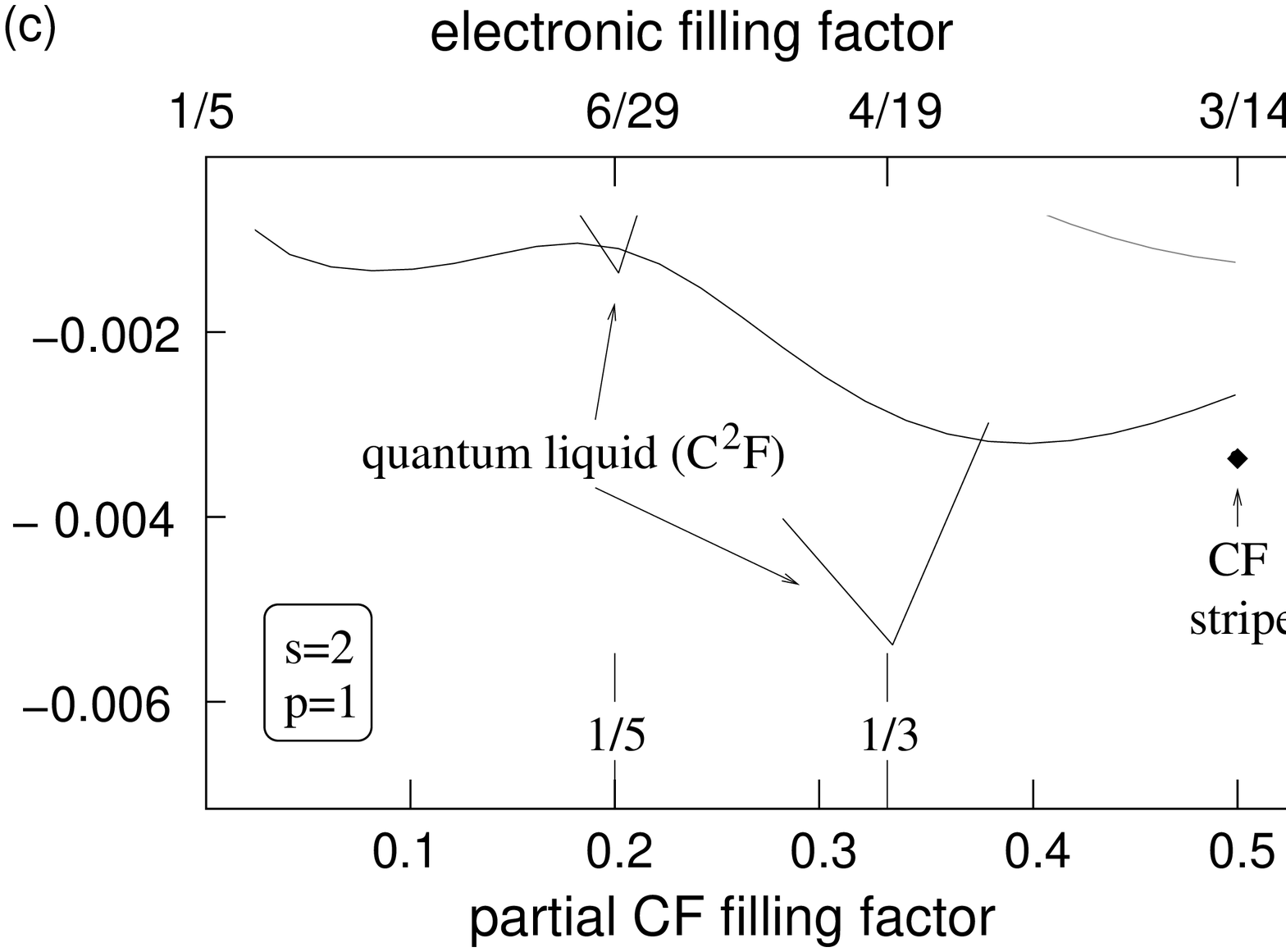}
\caption{Cohesive energies of the different CF phases for (a) $s=1$, $p=1$, 
(b) $s=1$, $p=2$, and (c) $s=2$, $p=1$ in units of $e^2/\epsilon l_B$.}
\label{fig01}
\end{figure}

The results for the cohesive energies of the different CF phases, given by
the Eqs.\ (\ref{equ03}), (\ref{equ04}), and (\ref{equ05}), are shown in
Fig.\,\ref{fig01} for three specific CF-LLs. Fig.\,\ref{fig01}(a) shows the 
results for CFs carrying $2$ flux quanta ($s=1$) in the first excited CF-LL 
($p=1$). The discussion is limited to the CF filling factor range 
$0<\nubar^*<1/2$, which corresponds to a range of the electronic filling factor
$1/3<\nu<3/8$. In analogy with the electronic case, the range $1/2<\nubar^*<1$ 
is related to the shown part by the particle-hole symmetry. Our energy 
calculations suggest that a spin-polarized Laughlin-type quantum liquid of CFs 
has a lower energy than the competing CF-solid phases around
$\nubar^*=1/3$ and $1/5$, which correspond to $\nu=4/11$ and $6/17$,
respectively. 
Especially the $4/11$ state is found to be largely favored, as expected from
the form of the pseudopotentials; one finds that $V_3(s=p=1)$ is smaller than 
its neighboring pseudopotentials \cite{goerbig04}. This stabilizes the 
$\stilde=1$ Laughlin state, which has its largest weight at the corresponding 
angular momentum and which screens $V_1$ completely \cite{haldane}. Indeed,
a spin-polarized FQHE at $\nu=4/11$ has been observed by Pan {\sl et al.}
\cite{pan}. The
CF-WC becomes stable at lower densities, $\nubar^*\lesssim 0.15$, as well as
at larger values $\nubar^*\gtrsim 0.4$. At $\nubar^*=1/2$, the CF-stripe phase
has the lowest energy, which may lead to the anisotropic longitudinal 
resistance at $\nu=13/8$, recently observed by Fischer {\sl et al.} 
\cite{fischer}. Contrary to the electronic case in the first excited LL 
\cite{goerbig03}, a CF-bubble crystal with two CFs ($M=2$) per lattice site 
never has a lower energy than the CF-WC ($M=1$). Although screening of the
CF interaction potential affects the activation gaps 
\cite{goerbig04} and may modify the transition points between the 
different phases, the investigated reentrance phenomenon persists in $1/3<\nu
<2/5$.
The effect of screening, finite sample width, and impurities on these phases 
will be discussed elsewhere \cite{goerbigUP}. 

For $s=1$ and $p=2$  
[Fig.\,\ref{fig01}(b)], {\sl i.e.} in the range $2/5<\nu<5/12$, the CF 
quantum liquid ceases to be the ground state at $\nubar^*=1/3$, where the 
CF-WC has the lowest energy. However, an incompressible CF liquid is found to 
be stable at $\nubar^*=1/5$. In contrast to the CF-LL $p=1$, a CF-bubble 
crystal with $M=2$ has a lower energy than the CF-WC around
half-filling, but a CF-stripe phase is found to be the ground state
at $\nubar^*=1/2$. 
For $1/5<\nubar^*<3/14$, which corresponds to $p=1$ for CFs carrying $4$ flux
quanta ($s=2$) [Fig.\,\ref{fig01}(c)], a $\nubar^*=1/3$ Laughlin state of
CFs is the lowest-energy state at $\nu=4/19$, whereas the liquid state is 
extremely close in energy to the CF-WC at $\nubar^*=1/5$. In the presence of
impurities, which lower the energy of the solid phases more significantly
than the quantum-liquid energies \cite{goerbig03}, a $\nubar^*=1/5$ 
Laughlin-type state might vanish. Impurities may also affect the $6/17$
state. In all cases an insulating CF-WC, which
is the natural ground state at lower densities, has a lower energy also at 
rather large filling factors, $\nubar\gtrsim0.4$. This may
lead to reentrance phenomena, which would be the CF analogue of the 
reentrant IQHE observed at lower magnetic fields \cite{exp3}.

The stability of C$^2$F states has been a controversial issue during the last
decade. Numerical investigations in the Haldane-Halperin hierarchy 
\cite{haldane,halperin} by B\'eran and Morf denied the stability of a 
spin-polarized $4/11$ state, whereas they found a small gap in the absence of
complete spin-polarization \cite{beran}. Their results have been
confirmed by W\'ojs and 
Quinn \cite{wojsquinn} and by numerical-diagonalization studies in 
the CF wave-function approach by Mandal and Jain \cite{mandaljain} and 
Chang {\sl et al.} \cite{chang}. However, more recent studies by Chang and Jain
\cite{changjain}, which hint to a stable $4/11$ state, contradict
their previous results. The size of the diagonalized
system remains too small to allow for a conclusive answer of its
stability in the thermodynamic limit.

In conclusion, we have calculated the energies of competing CF-solid and
liquid phases in a recently developed model of interacting CFs 
\cite{goerbig04}. Incompressible quantum liquids of spin-polarized CFs, 
which may be interpreted 
in terms of C$^2$Fs \cite{goerbig04,lopezfradkin}, are found to be stable 
at $\nu=4/11,6/17,4/19$, and $11/27$. Around half-filling of the topmost 
CF-LLs, a 
CF-stripe phase has the lowest energy, but may compete with a Pfaffian state
\cite{MR1}, which has been omitted in the present studies. The
fact that  
insulating CF-bubble crystals occur at different filling factors, which 
surround the CF-liquid phases, may lead to an experimentally observable 
reentrance of the FQHE in high-quality samples, in analogy to the reentrant 
IQHE in the first excited electronic LL \cite{exp3}.

We acknowledge fruitful discussions with R.\ Mani, R.\ Moessner, R.\ Morf,
and A.\ W\'ojs. 
This work was supported by the Swiss National Foundation for
Scientific Research under grant No.~620-62868.00.

\end{document}